\documentclass[letterpaper, 10 pt, conference]{ieeeconf}

\let\labelindent\relax               % to avoid conflicts with enumitem package

\IEEEoverridecommandlockouts % for the thanks command

\usepackage{amsthm}     % for the theorems environments
\usepackage{amssymb}    % for extra symbols
\usepackage{mathtools}  % for defining custom math symbols
\usepackage[all]{xy}    % for scripting graphs
\usepackage{enumitem}   % for list environments configuration
\usepackage{listings}   % for the lstlisting environment
\usepackage{framed}     % for frame environment
\usepackage{array}      % for column with configuration
\usepackage{comment}    % for the comment environment
\usepackage{etoolbox}   % for defining toggles
\usepackage{changes}    % for tracking changes
\usepackage{pgfplots}   % for drawing plots

\usepackage[caption=false,font=footnotesize]{subfig} % for subfigure configuration

\setlist[itemize]{             % itemize environment global configuration
  label=\scriptsize $\bullet$, % sets default itemize label
  topsep=2pt,                  % sets the separation from the top of the list
  itemsep=0pt,                 % sets no extra separation between items
} 

\setlist[enumerate]{           % enumerate environment global configuration
  topsep=2pt,                  % sets the separation from the top of the list
  itemsep=0pt,                 % sets no extra separation between items
} 

\newtoggle{examples}
\toggletrue{examples} % comment to exclude examples and insights

\renewcommand{\emph}[1]{\textit{#1}} % to avoid underlining

\renewcommand{\l}{\ell}

\newcommand{\D}{\rightarrow}

\newcommand{\ma}[1]{\tilde{#1}}
\renewcommand{\#}{\text{\rapprox}}
\newcommand{\rapprox}{\protect\raisebox{.09em}{\:\!\protect\reflectbox{\protect\rotatebox[origin=c]{90}{\ensuremath{\approx}}}}}

\newcommand{\B}{\rightsquigarrow}

\newcommand{\card}[1]{\left\vert{#1}\right\vert}

\renewcommand{\>}{\rangle}
\newcommand{\ldot}{\;.\;}

\newcommand{\always}{\text{\protect\raisebox{-.5pt}{\ensuremath{\square}}}}
\newcommand{\eventually}{\text{\protect\raisebox{.5pt}{\ensuremath{\lozenge}}}}

\newcommand{\step}[2]{\overset{#1}{\D}_{#2}}

\newcommand{\hop}[2]{\overset{#1}{\B}_{#2}}
\newcommand{\rnum}[1]{\uppercase\expandafter{\romannumeral #1\relax}}
\newcommand{\mk}[1]{\mathit{#1}}

\newtheoremstyle{thmstyle}
  {5pt} % Space above
  {5pt} % Space below
  {} % Body font
  {} % Indent amount
  {\bfseries} % Theorem head font
  {.} % Punctuation after theorem head
  {5pt} % Space after theorem head
  {} % Theorem head spec (can be left empty, meaning `normal')

\theoremstyle{thmstyle}

\newtheorem{definition}{Definition}
\newtheorem{notation}{Notation}
\newtheorem{property}{Property}
\newtheorem{lemma}{Lemma}

\newtheoremstyle{expstyle}
  {5pt} % Space above
  {5pt} % Space below
  {\slshape} % Body font
  {} % Indent amount
  {\bfseries} % Theorem head font
  {.} % Punctuation after theorem head
  {5pt} % Space after theorem head
  {} % Theorem head spec (can be left empty, meaning `normal')
  
\theoremstyle{expstyle}

\newtheorem{example}{Example}

\iftoggle{examples}{
  \newcommand{\insight}[1]{#1}
}{
  \excludecomment{example}
  \newcommand{\insight}[1]{}
}

\renewcommand{\k}[1]{\textit{#1}}
\renewcommand{\v}[1]{\texttt{#1}}

\newcolumntype{L}[1]{>{\raggedright\let\newline\\\arraybackslash\hspace{0pt}}m{#1}}
\newcolumntype{C}[1]{>{\centering\let\newline\\\arraybackslash\hspace{0pt}}m{#1}}
\newcolumntype{R}[1]{>{\raggedleft\let\newline\\\arraybackslash\hspace{0pt}}m{#1}}
\newcommand{\rb}[1]{\raisebox{-1.35pt}{#1}}

\newcommand{\node}[1]{*+[o][F]{#1}}
\newcommand{\vset}[2]{**[r]{\:\left\{\,\parbox{#1pt}{\ensuremath{#2}}\right\}}}

\begin{document}

%%%%%%%%%%%%%%%%%%%%%%%%%%%%%%%%%%%%%%%%%%%%%%%%%%%%%%%%%%%%%%%%%%%%%%%%%%%%%%%
%%  Title
%%%%%%%%%%%%%%%%%%%%%%%%%%%%%%%%%%%%%%%%%%%%%%%%%%%%%%%%%%%%%%%%%%%%%%%%%%%%%%%

\title{\vspace{-1cm} \LARGE \bf Technical Report: Directed Controller Synthesis of \\ Discrete Event Systems}

\author{Daniel~Ciolek$^1$, Victor~Braberman$^{1,3}$, Nicol\'as~D'Ippolito$^{1,2,3}$ and Sebasti\'an~Uchitel$^{1,2,3}$
\thanks{
$\!\!\!\!\!\!1.$ Departamento de Computaci\'on, Universidad de Buenos Aires, Argentina \protect\\
$2.$ Department of Computing, Imperial College, London, UK \protect\\
$3.$ CONICET}
\thanks{This work was submitted to the $55^{th}$ IEEE Conference on Decision and Control and was partially supported by grants ERC PBM-FIMBSE, ANPCYT PICT 2011-1774, ANPCYT PICT 2012-0724, ANPCYT PICT 2013-2341, UBACYT 20020130100384BA, UBACYT 20020130300036BA, CONICET PIP 112 201301 00688 CO, PIDDEF 04/ESP/15/BAA.}
}

\maketitle
\thispagestyle{empty}
\pagestyle{empty}

%%%%%%%%%%%%%%%%%%%%%%%%%%%%%%%%%%%%%%%%%%%%%%%%%%%%%%%%%%%%%%%%%%%%%%%%%%%%%%%
%%  Abstract
%%%%%%%%%%%%%%%%%%%%%%%%%%%%%%%%%%%%%%%%%%%%%%%%%%%%%%%%%%%%%%%%%%%%%%%%%%%%%%%

\begin{abstract}
This paper presents a Directed Controller Synthesis (DCS) technique for discrete event systems. The DCS method explores the solution space for reactive controllers guided by a domain-independent heuristic. The heuristic is derived from an efficient abstraction of the environment based on the componentized way in which complex environments are described. Then by building the composition of the components on-the-fly DCS obtains a solution by exploring a reduced portion of the state space. This work focuses on untimed discrete event systems with safety and co-safety (i.e. reachability) goals. An evaluation for the technique is presented comparing it to other well-known approaches to controller synthesis (based on symbolic representation and compositional analyses).
\end{abstract}

%%%%%%%%%%%%%%%%%%%%%%%%%%%%%%%%%%%%%%%%%%%%%%%%%%%%%%%%%%%%%%%%%%%%%%%%%%%%%%%
%%  Introduction
%%%%%%%%%%%%%%%%%%%%%%%%%%%%%%%%%%%%%%%%%%%%%%%%%%%%%%%%%%%%%%%%%%%%%%%%%%%%%%%

\section{Introduction} \label{sec:introduction}

Discrete Event Systems (DES) are dynamic systems that react to the occurrence of diverse discrete events. DES arise in many domains including robotics, logistics, manufacturing, and communication networks. These applications require control and coordination to ensure that the system goals are achieved. Numerous techniques for automatically verifying the correctness of controllers have been devised in the area of model checking. A different yet related approach, Controller Synthesis, pursues the automatic construction of controllers satisfying a formal specification.

The field of synthesis of controllers for DES was introduced by Ramadge and Wonham \cite{Ramadge:1987:SC} for controlling systems within a given set of constraints. In this setting, the environment (also called plant) and goals are specified using a formal language, and a procedure generates a correct by construction controller (or supervisor). The environment is usually modeled with state machines whose event sets are partitioned into controllable and uncontrollable actions. A controller must achieve its goals by dynamically disabling some of the controllable actions.

An alternative formulation of the controller synthesis problem considers the setting of $\omega$-regular languages \cite{Pnueli:1989:SRM}, instead of an automata-based DES. In this setting two-player games with Boolean objectives (where a win of one player coincides with a loss by the other player) have been suggested to reason about the interaction between the program and the environment \cite{Buchi:1990:SSC,Rabin:1968:DST}.

We address control problems for behavior models expressed as deterministic Labeled Transition Systems (LTS) and parallel composition ($\|$) defined broadly as a synchronous product. This setting allows to ease the modeling of complex environments by describing its behavior compositionally. Such descriptions are compact and, hence, obtaining the complete behavior of the environment via composition produces an exponential explosion. For this reason we want to avoid the computation of the composition ahead of time, building it on-the-fly instead (hopefully obtaining a solution by exploring a reduced portion of the state space).

Hereinafter we present the Directed Controller Synthesis (DCS) of DES for safety and co-safety (i.e. reachability) goals. Informally, we want the controller to \k{always} avoid safety violations while \k{eventually} ensuring co-safety objectives. More formally, we make use of the Linear Temporal Logic (LTL) \cite{Pnueli:1979:TSCP} operators. Thus, given the model of the environment components $E_0, \ldots, E_n$, the safety goal $G_S$ and the co-safety goal $G_C$, we look for a component $M$ such that when composed with the environment it satisfies the goals, namely $E_0\|\ldots\|E_n\|M \models \always \neg G_S \wedge \eventually G_C$.

DCS explores the solution space for reactive controllers on-the-fly guided by a domain-independent heuristic. The main contribution of this paper is the heuristic derived from an abstraction of the environment that exploits the componentized way in which complex environments are described. We highlight that componentization not only simplifies modeling, but also allows for preprocessing procedures that can improve the applicability of controller synthesis techniques.

The paper is organized as follows. Section~\ref{sec:related-work} comments on previous and related work. Section~\ref{sec:background} provides background. Section~\ref{sec:exploration} comments on details to take into account during the on-the-fly exploration of the state space. Section~\ref{sec:abstraction} presents the LTS abstraction and a heuristic derived from it. Section~\ref{sec:evaluation} reports on the results obtained with our implementation. Finally, section~\ref{sec:conclusions} discuses some conclusions and avenues for future work.

%%%%%%%%%%%%%%%%%%%%%%%%%%%%%%%%%%%%%%%%%%%%%%%%%%%%%%%%%%%%%%%%%%%%%%%%%%%%%%%
%%  Related Work
%%%%%%%%%%%%%%%%%%%%%%%%%%%%%%%%%%%%%%%%%%%%%%%%%%%%%%%%%%%%%%%%%%%%%%%%%%%%%%%

\section{Related Work} \label{sec:related-work}

Traditional controller synthesis techniques for safety objectives look for maximal (i.e. least restrictive) controllers \cite{Wonham:1987:SCS}, since safety properties can many times be trivially satisfied by restricting the system to an unproductive zone. Maximality comes at a cost in complexity given that it requires to explore the complete state space.

Monolithic approaches to controller synthesis work with the complete state space but, since it is exponential with respect to the size of the components, explicit representations are impractical. The Model Based Planner (MBP) presented in \cite{Gromyko:2006:TCS} uses Binary Decision Diagrams (BDD) \cite{Akers:1978:BDD} to keep a symbolic representation of the state space, in an attempt to cope with the state explosion problem. MBP is based on the NuSMV model checker and hence, its input is a model described in the SMV language, implementing different variations of goals (including liveness with CTL).

Contrarily to monolithic approaches, compositional approaches to controller synthesis (as the one implemented by Supremica \cite{Mohajerani:2011:CSDES}) build a controller for each component, such that when composed achieve the system requirements. Although this allows to greatly reduce the impact of the state explosion problem it also restricts its application to simple goals, such as safety, since it is not possible to guarantee composability for richer goals.

Controller synthesis is also related to the area of Automated Planning, a branch of artificial intelligence. However, the work in planning has been oriented mainly towards nonreactive environments, which are insufficient in the setting in which controller synthesis is applied. MBP provides a translation from a planning input to an SMV script and shows that a planning problem can be solved as a control problem by taking advantage of the advances on model checking \cite{Giunchiglia:2000:PMC}.

In planning the state space is usually represented explicitly, but the construction is performed on-the-fly guided by heuristics \cite{Bonet:2001:PHS}. Informed search procedures -- like A* \cite{Hart:1972:FBH} -- are used to perform a goal-directed exploration, generally obtaining a solution by inspecting a reduced portion of the state space. Many variations of the planning problem have been attacked this way and some 
\insight{-- especially those that contemplate nondeterminism such as FOND planning \cite{Muise:2014:NDP} --}
are closely related to controller synthesis.

Inspired by planning, informed search procedures have been introduced in model checking to accelerate the search for an error \cite{Edelkamp:2001:DEMC,Kupferschmid:2006:APHDM}. Although with this approach a trace to an error can be found faster, verifying the correctness of a model still requires to explore the complete state space. Despite the positive results of these techniques, their application to controller synthesis has been scarcely explored.

In \cite{Tripakis:1999:OCS} an on-the-fly synthesis method is presented for discrete and dense-time system, using depth-first search \emph{without a heuristic guide}. For linear hybrid systems, an on-the-fly informed search algorithm is presented in \cite{Alimguzhin:2013:OCSS}, which excels at design space exploration by quickly assessing whether a problem has no solution, \emph{rather than guiding the exploration towards a goal}. Finally the Controller Synthesis Module of the CIRCA architecture \cite{Goldman:2002:EIR}, which works on a timed automaton model, uses heuristics with a \emph{limited lookahead} and a verification procedure to detect and prune bad choices. In all these approaches the on-the-fly search procedure is similar to ours, but the setting differs since we consider as input the specification of the environment components, and hence \emph{the abstractions and heuristics diverge}.

Herein we present an extension to our previous work on the Modal Transition System Analyser (MTSA) \cite{D'Ippolito:2008:MTSA} tool that now includes the DCS algorithm. As mentioned before, in the present work we focus on safety and co-safety (i.e. reachability) goals, the former being optional while the latter mandatory since the heuristic needs a desirable end to properly direct the search.

%%%%%%%%%%%%%%%%%%%%%%%%%%%%%%%%%%%%%%%%%%%%%%%%%%%%%%%%%%%%%%%%%%%%%%%%%%%%%%%
%%  Background
%%%%%%%%%%%%%%%%%%%%%%%%%%%%%%%%%%%%%%%%%%%%%%%%%%%%%%%%%%%%%%%%%%%%%%%%%%%%%%%

\section{Background} \label{sec:background}

\begin{definition} [Labeled Transition Systems] \label{def:LTS}
A \k{Labeled Transition System} (LTS) is a tuple $E = (S, A, \D, s_0)$, where $S$ is a finite set of states, $A$ is its alphabet, $\D \: \subseteq (S \times A \times S)$ is a transition relation, and $s_0 \in S$ is the initial state.
\end{definition}

\begin{notation}
Let $E$ and $M$ be LTSs such that:
\begin{itemize}
\item $E = (S_E, A_E, \D_E, s_0)$, with $s$ and $s'$ states of $S_E$
\item $M = (T_M, A_M, \D_M, t_0)$, with $t$ and $t'$ states of $T_M$
\end{itemize}
\end{notation}

\begin{notation}[Step] \label{not:step}
We denote $(s,\l,s') \in \: \D$ by $s \step{\l}{} s'$.
\end{notation}

\begin{definition} [Parallel Composition] \label{def:parcomp}
The \k{Parallel Composition} $(\|)$ is a symmetric operator such that it yields an LTS $E \| M = (S_E {\times} T_M, A_E {\cup} A_M, \D_{E\|M}, \<s_0,t_0\>)$, where $\D_{E\|M}$ is a relation that satisfies the following rules:

\begin{small}
\begin{minipage}{0.45\linewidth}
\[\!\!\!\!\!\! \frac{s \step{\l}{E} s'}{\<s,t\> {\step{\l}{E\|M}} \<s',t\> }{\; {\scriptstyle \l \in (A_E {\setminus} A_M)}} \]
\end{minipage} $\;$
\begin{minipage}{0.45\linewidth}
\[ \frac{t \step{\l}{M} t'}{\<s,t\> {\step{\l}{E\|M}} \<s,t'\>}{\; {\scriptstyle \l \in (A_M {\setminus} A_E)}} \]
\end{minipage} \\[5pt]
\begin{minipage}{1.05\linewidth}
\[ \frac{s \step{\l}{E} s', t \step{\l}{M} t'}{\<s,t\> \step{\l}{E\|M} \<s',t'\>}{\; {\scriptstyle \l \in (A_E {\cap} A_M)}} \]
\end{minipage}
\end{small}

\vspace{10pt}

\end{definition}

\begin{example}
In Fig.\ref{fig:parcomp-example} we show an example of the application of the parallel composition operator ($\|$). Note that synchronizing actions produce an update in the states of both LTSs, while nonshared actions produce an update in only one LTS at a time. Observe also that shared actions that are not available from both LTSs at a given point cannot be executed (e.g. $d$ available from $t_0$ but not from $s_0$ cannot be executed from the initial state $\<s_0,t_0\>$).

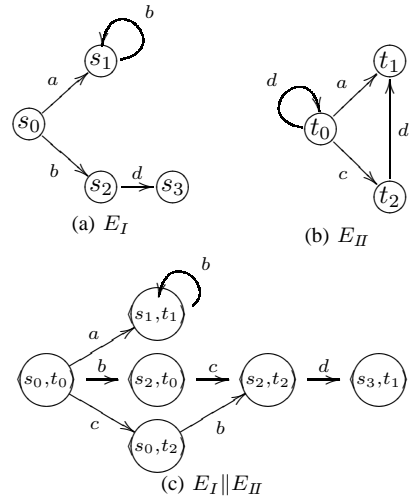
\begin{figure}[ht]
\centering
\subfloat[$E_I$]{
$\xymatrix@-1pc{
& \node{s_1} \ar_{b}@(r,u)[]         \\
\node{s_0} \ar^{a}[ur] \ar_{b}[dr]   \\
& \node{s_2} \ar^{d}[r] & \node{s_3} \\
}$
\label{fig:E0}
} \hspace{.5cm}
\subfloat[$E_{I\;\!\!\!I}$]{
$\xymatrix@-1pc{
& \node{t_1}                       \\
\node{t_0} \ar^{a}[ur] \ar_{c}[dr] \ar^{d}@(l,u)[] \\
& \node{t_2} \ar_{d}[uu]           \\
}$
\label{fig:M0}
} \hspace{.5cm}
\subfloat[$E_{I}\|E_{I\;\!\!\!I}$]{
$\xymatrix@-1pc{
& \node{_{\<s_1,t_1\>}} \ar_{b}@(r,u)[] \\
\node{_{\<s_0,t_0\>}} \ar^{a}[ur] \ar^b[r] \ar_{c}[dr] & \node{_{\<s_2,t_0\>}} \ar^{c}[r] & \node{_{\<s_2,t_2\>}} \ar^{d}[r] & \node{_{\<s_3,t_1\>}} \\
& \node{_{\<s_0,t_2\>}} \ar_{b}[ur] \\
}$
\label{fig:parcomp}
}
\caption{Parallel Composition Example}
\label{fig:parcomp-example}
\end{figure}
\end{example}

\insight{Given a partition of the alphabet in controllable and uncontrollable actions, a controller $M$ must achieve its goal $G$ by disabling some of the controllable actions in $E$, while only monitoring uncontrollable actions. }We say that a controller $M$ satisfies a goal $G$ in an environment $E$ if and only if every trace accepted by $E\|M$ satisfies $G$.

\begin{definition} [Trace] \label{def:trace}
A trace of an LTS $E$ is a sequence of labels $\pi {=} \l_0, \l_1 \ldots$ of $A_E$, for which there exists a sequence of states $s_0, s_1, \ldots$ of $S_E$ such that $\forall i {\geq} 0 \ldot s_i \step{\l_i}{E} s_{i+1}$.
\end{definition}

\begin{definition}[Safety] \label{def:safety}
We say that a trace $\pi$ satisfies a \k{safety} goal $G_S$ given as a set of labels, if and only if a label of $G_S$ never occurs in $\pi$. %In other words, the actions in $G_S$ are never executed.
\end{definition}

\begin{definition}[Co-Safety] \label{def:co-safety}
We say that a trace $\pi$ satisfies a \k{co-safety} goal $G_C$ given as a set of labels, if and only if there is at least one occurrence of a label of $G_C$ in $\pi$. %Or in other words, at least one action in $G_C$ is guaranteed to be executed.
\end{definition}

\begin{example}
In Fig.\ref{fig:controller-example} we show two controllers for $E_{I}\|E_{I\;\!\!\!I}$ that are guaranteed to reach $\{d\}$ assuming that only $a$, $b$ and $c$ are controllable. The controller in Fig.\ref{fig:maximal-controller-example} simply disables $a$ from the initial state to avoid entering a state where $d$ is no longer reachable. Note that this controller does not force the order of actions $b$ and $c$, that is, it imposes the least restrictive constraints. On the other hand, the controller in Fig.\ref{fig:non-maximal-controller-example} forces the order between actions $b$ and $c$, hence, despite being correct it is not a maximal controller.

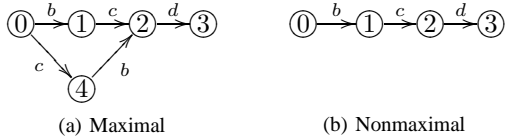
\begin{figure}[h]
\centering
\subfloat[Maximal]{
$\xymatrix@-1pc{
\node{0} \ar^b[r] \ar_{c}[dr] & \node{1} \ar^{c}[r] & \node{2} \ar^{d}[r] & \node{3} \\
& \node{4} \ar_{b}[ur] \\
}$
\label{fig:maximal-controller-example}
} \hspace{.5cm}
\subfloat[Nonmaximal]{
$\xymatrix@-1pc{
\node{0} \ar^b[r] & \node{1} \ar^{c}[r] & \node{2} \ar^{d}[r] & \node{3} \\
{\phantom{X}} \\
}$
\label{fig:non-maximal-controller-example}
}
\caption{Controllers for $E_{I}\|E_{I\;\!\!\!I}$ reaching $\{d\}$}
\label{fig:controller-example}
\end{figure}
\end{example}

\begin{notation} \label{not:set-step}
States of the composition are ordered pairs, however we may need to refer to the \k{step} relation for any ordering. For this purpose we may use the following notation:
\begin{small}
\[
\arraycolsep=1.5pt
\begin{array}{ll}
\{s,t\} {\step{\l}{E\|M}} \{s',t'\} \Leftrightarrow & \!
  \<s,t\> {\step{\l}{E\|M}} \<s',t'\> \vee
  \<s,t\> {\step{\l}{E\|M}} \<t',s'\> \vee \\ & \!
  \<t,s\> {\step{\l}{E\|M}} \<s',t'\> \vee
  \<t,s\> {\step{\l}{E\|M}} \<t',s'\>
\end{array}
\]
\end{small}

When using the set notation we relax the convention that $s$ and $s'$ belong to $E$ while $t$ and $t'$ belong to $M$ since the order of the states is made irrelevant.
\end{notation}

\begin{notation} \label{not:set-state}
For brevity we denote that a set of states $s_0, s_1, \ldots, s_n$ belongs to a set $q$, that is to say that $\forall i \ldot 0 \leq i \leq n \Rightarrow s_i \in q$, as $[s_1,s_2,\ldots,s_n]_q$.
\end{notation}

%%%%%%%%%%%%%%%%%%%%%%%%%%%%%%%%%%%%%%%%%%%%%%%%%%%%%%%%%%%%%%%%%%%%%%%%%%%%%%%
%%  On-the-fly Exploration
%%%%%%%%%%%%%%%%%%%%%%%%%%%%%%%%%%%%%%%%%%%%%%%%%%%%%%%%%%%%%%%%%%%%%%%%%%%%%%%

\section{On-the-fly Exploration} \label{sec:exploration}

In this section we discuss the main algorithm that constitutes DCS. Given a heuristic function that estimates the distance from a state to a goal, DCS looks for a controller by computing the parallel composition on-the-fly guided by the heuristic. The procedure is a modification of Best First Search (a classical informed search procedure) adapted to account for uncontrollable actions. We keep a priority queue of $\mk{open}$ states, ordered by their estimated distance to a goal, initialized only with the initial state. At each iteration the algorithm takes the most promising state from the queue and expands a child state following its best unexplored action. The expanded state is evaluated using the heuristic, and it is then inserted in the $\mk{open}$ queue (the value of the state is set as the estimate of its best unexplored action).

For controllable states we need only one successful successor, while for uncontrollable states all their successors have to be able to reach a goal. Thus we allow a controllable state to remain in the $\mk{open}$ queue as long as it has unexplored actions to broaden the search (i.e. competing with its descendants). Mixed states, that is states with both controllable and uncontrollable actions, are treated as uncontrollable states. This is because mixed states represent a race condition between the controller and the environment, which can always be won by the environment. Thereby the validity of the controller cannot depend on the result of race conditions. This treatment of controllable and uncontrollable nodes is similar to \textsc{and/or} search procedures \cite{Simon:1971:OSAOG}.

When a state is recognized either as a co-safety goal or a safety error, it is marked and this information is propagated back to its ancestors until an interrupting state is reached and reopened. The propagation of an error is interrupted when a controllable ancestor with unexplored actions is reached. Whereas, the propagation of a goal is interrupted when an uncontrollable ancestor with unexplored actions is reached. The process is repeated until the initial state is marked as a goal or an error. In the former the states reachable from the initial state form a controller, while in the latter there is no controller.

Observe that while there is an obvious advantage in exploring the most promising controllable action first, there is no such advantage in exploring the best uncontrollable action since all such actions must lead to a goal. Thus we could avoid the computation of the heuristic for uncontrollable states. Instead we use the reverse ranking and expand the least desirable uncontrollable action first. We do this in an attempt to find an error as fast as possible and consequently being able to close the state avoiding futile exploration.

In the worst case, the heuristic misguides the search and the algorithm explores the complete state space. With the additional cost of computing the heuristic for each state, the complexity of the algorithm is worse than its monolithic counterpart. That is, DCS could take an exponential amount of time. In spite of that, if the heuristic accurately guides the exploration with a small computational overhead, great savings could be obtained.

% However, the abstraction is strongly related to the real environment and these cases are near to impossible to come by in a real application.

%%%%%%%%%%%%%%%%%%%%%%%%%%%%%%%%%%%%%%%%%%%%%%%%%%%%%%%%%%%%%%%%%%%%%%%%%%%%%%%
%%  Abstraction and Heuristic
%%%%%%%%%%%%%%%%%%%%%%%%%%%%%%%%%%%%%%%%%%%%%%%%%%%%%%%%%%%%%%%%%%%%%%%%%%%%%%%

\section{Abstraction and Heuristic} \label{sec:abstraction}

In this section we present the abstraction we use and the heuristic it induces. The main goal of the heuristic is to provide an estimate of the distance from a state to a goal without computing the parallel composition. However, in order to provide informative estimates the effects of synchronization need to be taken into account to some point. For this reason we build an abstraction of the environment that, instead of considering the cross product of states, it works on sets of states. More precisely, if the states contained in the set can synchronize (with the standard parallel composition), then the synchronizing action must be available from the set.

At first glance considering the power set of states may seem detrimental, but we also apply a rule of monotonic growth. That is, the reachable sets of states are restricted to sets that contain all the traversed states from the initial state. Thus the abstraction size is polynomial with respect to the number of states in the components. We then use the length of the paths that reach a co-safety goal in the abstraction as an estimate of the real distance to the goal.

\subsection{LTS Abstraction and Composition}

\insight{Intuitively, we build an abstraction that, once it reaches a point where a state is covered, it never drops a state. The abstraction behaves as if, after a transition, not only a new state is covered but also the source state is not left. This progressively distances the abstraction from the real environment (as it collects states), yet it keeps information about the causal relationship between actions.}

We begin by defining the \k{LTS abstraction} for a single LTS, that yields an LTS whose reachable region forms \emph{a sequence of sets of states}.

\begin{definition} [LTS Abstraction] \label{def:ma}
The abstraction of an LTS $E$ is an LTS $\ma{E} = (2^{S_E}, 2^{A_E}, \B_E, \{s_0\})$, where $\B_E$ is the relation that satisfies:
\vspace{-.1cm}
\begin{small}
\[
\arraycolsep=1pt
\begin{array}{ll}
q {\hop{A}{E}} q' \Leftrightarrow &
  \big(\forall s,s',a \ldot [s]_{q} \wedge s {\step{a}{E}} s' \Rightarrow [s,s']_{q'} \wedge a {\in} A\big) \wedge \\ &
  \big(\forall s' \ldot [s']_{q'} \Rightarrow [s']_{q} \vee \exists s,a \ldot [s]_{q} \wedge a {\in} A \wedge s {\step{a}{E}} s'\big)
\end{array}
\]
\end{small}
\end{definition}

Observe that in the abstraction not only the states are sets of the original states, but also the labels are sets of the original labels. Furthermore, there is only one set of actions available from each set of states, that is, the maximal set of enabled actions.

We now define the \k{abstracting composition}, an operation for composing these abstracted LTSs. Informally, this operation works by applying a relaxed synchronization rule that follows that of ($\|$), but considering sets of states.

\begin{definition} [Abstracting Composition] \label{def:mac}
The \k{Abstracting Composition} $(\#)$ is a symmetric operator in the spirit of $(\|)$ that yields an abstract LTS $\ma{E} \# \ma{M} = (S_{\ma{E}} \cup S_{\ma{M}}, A_{\ma{E}} \cup A_{\ma{M}}, \B_{\ma{E}\#\ma{M}}, \ma{m}_0 \cup \ma{t}_0\})$ and $\B_{\ma{E}\#\ma{M}}$ is the relation that satisfies:
\vspace{-.1cm}
\begin{small}
\[
\arraycolsep=0pt
\begin{array}{l}
q \hop{A}{\ma{E}\#\ma{M}} \!q' \Leftrightarrow \\
\quad\; \big(\forall s,\!s'\!\!,\!t,\!t'\!\!,\!a {\ldot} [s,\!t]_{q} {\wedge} \<s,\!t\> {\step{a}{E\|M}} \<s'\!\!,\!t'\> \Rightarrow [s'\!\!,\!s,\!t'\!\!,\!t]_{q'} {\wedge} a {\in} A\big) \wedge \\
\quad\; \big(\forall s' {\ldot} [s']_{q'} \Rightarrow [s']_{q} {\vee} \exists s,\!t,\!t'\!\!,\!a \ldot [s,t]_{q} {\wedge} a {\in} A {\wedge} \{\!s,\!t\!\} {\step{a}{E\|M}} \{\!s'\!\!,t'\!\} \big)
\end{array}
\]
\end{small}
\end{definition}

Note that we do not need to compute $E\|M$ since we only check locally that states $s$ and $t$ synchronize following the rules in Def.\ref{def:parcomp}.

\begin{property} \label{prt:reach-length}
Given LTSs $E$ and $M$ the number of sets of states reachable in $\ma{E}\#\ma{M}$ is, in worst case, $\card{S_E} + \card{S_M}$.

\begin{proof}
The proof is immediate given that once every state in $S_M \cup S_E$ is included in a set of states $q$, no other set of states can be reached. Thus, in the worst case each step contributes only one fresh state, leading to $\card{S_E} + \card{S_M}$ steps before reaching the last possible fresh state.
\end{proof}
\end{property}

\begin{example}
In Fig.\ref{fig:abstraction-examples} we show examples for the \k{LTS abstraction} (Def.~\ref{def:ma}) and the \k{abstracting composition} (Def.~\ref{def:mac}) of LTSs $E_{I}$ and $E_{I\;\!\!\!I}$. In Fig.\ref{fig:ma-example} the abstraction of $E_{I}$ is depicted. The first step of $\ma{E}_{I}$ is constituted by the actions $a$ and $b$ (the actions available from $s_0$), reaching a set of states with $s_1$ (the target of $a$) and $s_2$ (the target of $b$), plus states already reached in previous steps (i.e. $s_0$). The second step only adds the action $d$ available from $s_2$ reaching $s_3$, once all the reachable states are included in the set a self loop captures potential infinite behaviours.

In Fig.\ref{fig:mac-example} a similar construction is shown for the abstracting composition between $\ma{E}_{I}$ and $\ma{E}_{I\;\!\!\!I}$. Note that despite the fact that $d$ is available from $t_0$ it is not available from $\{s_0,t_0\}$ since it is a synchronizing action not enabled at $s_0$.
\vspace{-.25cm}
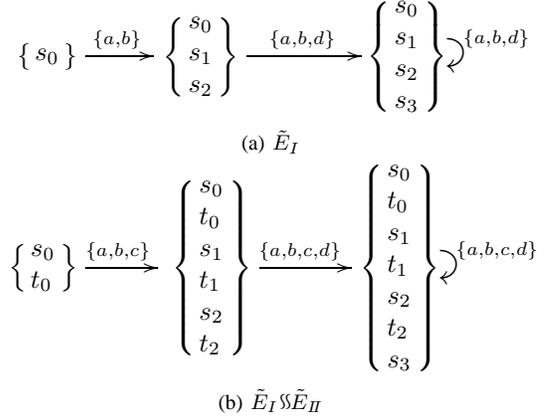
\begin{figure}[ht]
\centering
\subfloat[$\ma{E}_{I}$]{
$\xymatrix@+0.7pc{
\vset{11}{s_0} \ar^<<<<<{\{a,b\}}[r(0.75)] & \vset{11}{s_0\\s_1\\s_2} \ar^<<<<<<<<{\{a,b,d\}}[r(0.75)] & \vset{11}{s_0\\s_1\\s_2\\s_3} \rotatebox[origin=c]{-90}{\Large \ensuremath{\curvearrowright}}^{\{a,b,d\}} \\
}$
\label{fig:ma-example}
} \hspace{.15cm}
\subfloat[$\ma{E}_{I}\#\ma{E}_{I\;\!\!\!I}$]{
$\xymatrix@+0.7pc{
\vset{11}{s_0\\t_0} \ar^{\{a,b,c\}}[r] & \vset{11}{s_0\\t_0\\s_1\\t_1\\s_2\\t_2} \ar^<<<<<<{\{a,b,c,d\}}[r(0.6)] & \vset{11}{s_0\\t_0\\s_1\\t_1\\s_2\\t_2\\s_3} \rotatebox[origin=c]{-90}{\Large \ensuremath{\curvearrowright}}^{\{a,b,c,d\}} \\
}$
\label{fig:mac-example}
}
\caption{LTS abstraction and composition examples}
\label{fig:abstraction-examples}
\end{figure}
\vspace{-.25cm}
\end{example}

Despite looking very dissimilar there is a strong relation between a model and its abstraction. In particular if $\pi$ is a trace of $E\|M$, $\pi$ is a path contained in $\ma{E}\#\ma{M}$.
%\begin{definition} [Contained Path] \label{def:c-path}
We say that a sequence of actions $\pi {=} \l_0,\l_1 \ldots$ is a \k{contained path} in $\ma{E}\#\ma{M}$ if and only if there is a trace $\Pi {=} L_0,L_1 \ldots$ of $\ma{E}\#\ma{M}$ such that $\forall i \ldot \l_i \in L_i$.
%\end{definition}
Moreover, we may abuse notation and denote that for a given label $\l_i$ exists a set $L_i$ such that $\l_i \in L_i \wedge q \hop{L_i}{\ma{E}\#\ma{M}} q'$ using directly $\l_i$ as the label of the transition as follows: $q \hop{\l_i}{\ma{E}\#\ma{M}} q'$.

Note that by Def.~\ref{def:mac} every trace of $E\|M$ is a path contained in $\ma{E}\#\ma{M}$. However, not every contained path can be mapped to a trace in $E\|M$. Therefore after computing the abstraction some paths will relate to traces while others will not, yet a priori we have no way to distinguish between these cases. We can easily filter some paths that are obviously uninteresting\insight{, such as the paths that contain noncomponent-steps or non-synchronizing-steps (e.g. $s_0 {\step{c}{\ma{E}_{I}\#\ma{E}_{I\;\!\!\!I}}} t_2$)}. For this reason we define \k{abstracting path} to more precisely capture the notion of path that will help us estimate the distance to a goal.

\begin{definition} [Abstracting Path] \label{def:a-path}
We say that a sequence of actions $\l_0, \l_1, \ldots$ is an \k{abstracting path} of $\ma{E}\#\ma{M}$ if and only if there is a sequence of states $s_0, s_1, \ldots$ of $S_E \cup S_M$ and a sequence of sets of states $q_0, q_1, \ldots$ of $S_{\ma{E}\#\ma{M}}$ such that:
\vspace{-.1cm}
\begin{small}
\[
\arraycolsep=1pt
\begin{array}{ll}
\forall i \ldot & ([s_i]_{q_i} \wedge q_i \hop{\l_i}{\ma{E}\#\ma{M}} q_{i+1} \wedge \exists t,\!t'\! \ldot \{s_i,t\} \step{\l_i}{E\|M} \{s_{i+1},t'\}) \vee \\ &
(s_i {=} s_{i+1} \wedge \l_i {=} \tau)
\end{array}
\]
\end{small}
\end{definition}

This definition allows not only for valid intra-LTS steps \insight{(e.g. $s_0 {\step{a}{\ma{E}_{I}\#\ma{E}_{I\;\!\!\!I}}} s_1$)}, but it also allows for inter-LTS steps as long as they are potential synchronization steps \insight{(e.g. $s_0 {\step{a}{\ma{E}_{I}\#\ma{E}_{I\;\!\!\!I}}} t_1$)}.
\insight{This is important since an isolated component may not be able to reach a goal, but it may be necessary for an intermediate synchronization step.}
Observe that we add a special action $\tau$ to ``delay'' a step from a state $[s_i]_{q_i}$ to actions available from $[s_i]_{q_{i+1}}$. This allows to consider paths that skip necessary synchronization steps, while still considering them in the estimation. This definition filters the paths that are not solely constituted by these intra-LTS and inter-LTS steps. However, it remains that not every abstracting path is a trace,
\insight{
yet we use the length of these paths as an estimate of the expected distance to a co-safety goal.}

\begin{property} \label{prt:a-path-trace}
Given LTSs $E$ and $M$ every trace $\pi$ of $E\|M$ is an \k{abstracting path} of $\ma{E}\#\ma{M}$.

\begin{proof}
The proof is straightforward, since given that $\pi = \l_0, \l_1, \ldots$ is a trace of $E\|M$ there exists a sequence of states:
\vspace{-.1cm}
\begin{small}
\[
\<s_0,t_0\> \step{\l_0}{E\|M} \<s_1,t_1\> \step{\l_1}{E\|M} \ldots
\]
\end{small}
Then by Def.~\ref{def:mac} there exists a sequence of sets of states $q_0, q_1, \ldots$ such that $\forall i \ldot [s_i,t_i]_{q_i}$ and
\vspace{-.1cm}
\begin{small}
\[
q_0 \hop{\l_0}{\ma{E}\#\ma{M}} q_1 \hop{\l_1}{\ma{E}\#\ma{M}} \ldots
\]
\end{small}
Therefore by Def.~\ref{def:a-path} $\pi$ is an \k{abstracting path} of $\ma{E}\#\ma{M}$.
\end{proof}
\end{property}

We compute the estimate simply by taking the length of an \k{abstracting path} that reaches a co-safety goal. Paths that reach a safety violation are considered to have an infinite ($\infty$) distance to the goal. Interestingly, this can never overestimate the distance to the goal, but it can underestimate it. Heuristics that do not overestimate the distance to the goal are called admissible and have been studied in the literature since they enjoy useful properties (e.g. early error detection).

\begin{lemma}[Admissibility] \label{lem:admissibility}
The length of an \k{abstracting path} that reaches a co-safety goal is an admissible heuristic, that is, it does not overestimate the distance to the goal.

\begin{proof}
Note that in the worst case a path $\pi$ reaching a goal is a trace of the environment, thus its length estimates exactly the distance to the goal. In any other case $\pi$ is not a valid sequence of actions of the environment, that is to say is an artifact of the weakened composition rules of the abstraction. Hence $\pi$ may skip some necessary intermediate steps required to actually reach the goal in the environment. Therefore, the length of $\pi$ underestimates the distance to the goal.
\end{proof}
\end{lemma}

%%%%%%%%%%%%%%%%%%%%%%%%%%%%%%%%%%%%%%%%%%%%%%%%%%%%%%%%%%%%%%%%%%%%%%%%%%%%%%%
%%  Implementation Details
%%%%%%%%%%%%%%%%%%%%%%%%%%%%%%%%%%%%%%%%%%%%%%%%%%%%%%%%%%%%%%%%%%%%%%%%%%%%%%%

\subsection{Heuristic Computation} \label{sec:implementation}

In this section we show how we build the abstraction and compute the heuristic estimates for the actions available from a given state. For computing the heuristic we need all the \k{abstracting paths} reaching a goal or an error from an initial state. For this we build a graph whose paths are all such \k{abstracting paths}, which is naturally induced from Def.~\ref{def:a-path}. However, instead of explicitly representing the special action $\tau$, we represent the ``distance'' between states by a positive integer that we call \k{generation}. 

\begin{definition}[Generation] \label{def:generation}
Given an abstracted LTS $\ma{E}$ we define the \k{generation} of a state $s$ of $S_{\ma{E}}$ as the index $g$ of the \emph{sequence of sets of states} of $\ma{E}$ in which the state $s$ appears for the first time, denoted $s^g$. More formally,
\vspace{-.1cm}
\begin{small}
\[
s^g \Leftrightarrow g = \max\limits_i \, \{ \, i \mid \forall \, 0 \leq j < i, \exists \l_j \ldot q_j \hop{\l_j}{\ma{E}} q_{j+1} \wedge s \not\in q_j \}
\]
\end{small}
\end{definition}

\insight{The \k{generation} is useful since given a transition from a state $s_i$ into $s_j$ with generations $n$ and $m$ respectively, we can deduce the ``distance'' between them by subtracting $|m - n|$.}

\begin{definition}[Abstracting Path Graph] \label{def:a-path-graph}
Given LTSs $E$ and $M$, the \k{Abstracting Path Graph} of $\ma{E}\#\ma{M}$ is a graph $\mathcal{G}(\ma{E}\#\ma{M}) = (\mathcal{V},\mathcal{E},\mk{gen})$ where:
\begin{itemize}
\item $\mathcal{V} = S_E \cup S_M$, is a set of vertices formed by states
\item $\mathcal{E} = \{ (s_i,\l,s_j) \mid \exists q,q',t,t' \ldot [s_i,t]_{q} \wedge q \hop{\l}{\ma{E}\#\ma{M}} q' \wedge \{s_i,t\} \step{\l}{E\|M} \{s_j,t'\} \}$, is a set of labeled edges
\item $\mk{gen} : V \rightarrow \mathbb{N}$, is a function mapping states to their respective \k{generations}
\end{itemize}
\insight{Informally, two vertices of the graph are connected by an edge if and only if there is a step, connecting the corresponding states, that is part of an \k{abstracting path}.}
\end{definition}

\begin{example}
In Fig.\ref{fig:a-path-graph-example} we show the \k{abstracting path graph} of $\ma{E}_{I}\#\ma{E}_{I\;\!\!\!I}$. Each edge of the graph represents a possible step in the abstraction that is a part of an abstracting path. The first column of the graph contains the states at generation $0$ ($s_0$ and $t_0$), the second column contains the fresh states at generation $1$ ($s_1$, $s_2$, $t_1$ and $t_2$), while the third column contains the fresh states at generation $3$ ($s_3$). An edge that skips one or more columns (or generations) represents a fragment of an \k{abstracting path} with $\tau$-steps. For example, the edge labeled $d$ connecting $t_0$ with $s_3$ represents a $\tau$-step followed by action $d$. This is because $d$ is not available from the initial state and is delayed (i.e. the path has a length of $2$).
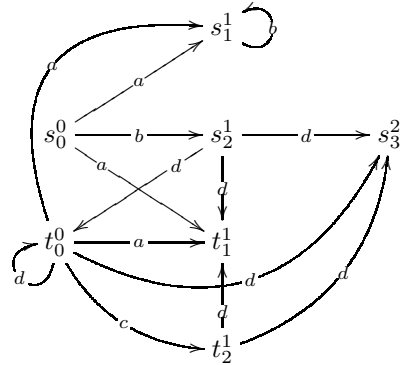
\begin{figure}[ht]
\centering
$
\xymatrixcolsep{4pc}
\xymatrix@+0pc{
& s_1^1 \ar|-{\,b\,}@(rd,ru)[] \\
s_0^0 \ar|-{\,a\,}[ur] \ar|-{\,b\,}[r] \ar|-<<<<{\,a\,}[dr] & s_2^1 \ar|-{\,d\,}[r] \ar|-{\,d\,}[d] \ar|-<<<<{\,d\,}[dl] & s_3^2 \\
t_0^0 \ar|-{\,a\,}[r] \ar|-{\,c\,}@/_1pc/[dr] \ar|-{\,d\,}@(d,l)[] \ar|-{\,a\,}@/^3.5pc/[uur] \ar|-{\,d\,}@/_3pc/[urr] & t_1^1 \\
& t_2^1 \ar|-{\,d\,}@/_1.5pc/[uur] \ar|-<<{\,d\,}[u]
}$
\caption{Abstracting Path Graph $\mathcal{G}(\ma{E}_{I}\#\ma{E}_{I\;\!\!\!I})$} \label{fig:a-path-graph-example}
\end{figure}

We can use this graph to build a ranking of the available actions from the initial state of $E_{I}\|E_{I\;\!\!\!I}$, namely $\<s_0,t_0\>$, by measuring the distance to the co-safety goal $\{d\}$. The available actions are $a$, $b$ and $c$ (as mentioned before $d$ is disabled from the initial state). The shortest paths to $d$ starting with $b$ and $c$ have a length of $2$, while there are no paths reaching $d$ starting with $a$. Hence, we estimate an infinite distance to the goal for every path starting with $a$. Ranking the actions by these estimates directs us to explore the actions $b$ and $c$. In spite of that, we are underestimating the distance to the goal since our estimation assumes that immediately after either $b$ or $c$ we would be able to execute $d$. This is an artifact of the weak composition rules since both $b$ and $c$ need to be executed (in any order) before $d$. Still, these estimates effectively guide the exploration towards the goal.
\end{example}

In Fig.~\ref{lst:build-abstraction} we present a procedure that given a state $\<s_{E_0},\ldots,s_{E_n}\>$ of the composition $E_0 \| \ldots \| E_n$, and sets of actions desired to $\mk{reach}$ and to $\mk{avoid}$, it returns the information required to compute a heuristic ranking:

\begin{enumerate}
\item $\mk{errors}$, a set of states that do not conduce to a goal.
\item $\mk{goals}$, a set of transitions connected by a $\mk{reach}$ action.
\item $\mk{edges}$ of the \k{abstracting path graph}.
\item $\mk{generations}$ of the states in the graph.
\end{enumerate}

The procedure works by iteratively building the \emph{sequence of sets of states} produced by the abstraction (Def.~\ref{def:mac}) and, at the same time, the \k{abstracting path graph} (Def.~\ref{def:a-path-graph}). At each iteration a new set $q_g$ is considered, where $g$ indicates the current \k{generation}. We then compute the set of $\mk{ready}$ transitions, that is, the steps available from $q_g$ following the relaxed synchronization rule from Def.~\ref{def:mac}. Observe that once a transition is processed it is never considered again. Also note that states are considered errors until they show potential to reach a goal (i.e. deadlock states are treated as errors). Since the procedure processes each transition once its complexity is $\Theta(\card{\B_{\ma{E}_0\#\ldots\#\ma{E}_n}})$,
\insight{
which considering the weakened synchronization rules, can be bigger than the actual transitions in the individual components.}
In the worst case, the complexity cannot surpass the connection of every state by every action, that is to say that its upper-bound complexity is $O\big((\sum_{i=0}^{n}{\card{A_{E_i}}})(\sum_{i=0}^{n}{\card{S_{E_i}}})^2\big)$.

\begin{figure}
\vspace{0.2cm}
\begin{framed}
\begin{lstlisting}
buildAbstraction($\,\<s_{E_0}, \ldots, s_{E_n}\>, reach, avoid\,$)
  $errors = q_0 = \overset{n}{\underset{i=0}{\bigcup}} s_i$
  $goals = processed = edges = generations = \emptyset$
  for each $s \in q_0$ do $generations[s] = 0$
  $ready = \{ \, (s,\l,s') \mid \exists q' \!\!\ldot q_0 {\hop{\l}{\ma{E}_0\#...\#\ma{E}_n}} q' \wedge (s,\l,s') {\not\in} edges \}$
  $g = 1$
  while $ready \neq \emptyset$ do
    for each $(s,\l,s') \in ready$ do
      $errors = errors \setminus \{s\}$
      $processed = processed \cup \{ s \}$
      $edges = edges \cup \{ (s,\l,s') \}$
      if $\l \in reach$ then
        $goals = goals \cup \{ (s,\l,s') \}$
      else if $s' \not\in processed$ then
        $errors = errors \cup \{ s' \}$
        $generations[s'] = g$
        if $\l \not\in avoid$ then
          $q_{g+1} = q_g \cup \{ s' \}$
    $ready = \{ (s,\!\l,\!s') \!\!\mid \!\exists q' \!\!\ldot q_{g+1} {\hop{\l}{\ma{E}_0\#...\#\ma{E}_n}} q' \!\wedge (s,\l,s') {\not\in} edges \}$
    $g = g + 1$
  return $\<errors, goals, edges, generations\>$
\end{lstlisting}
\end{framed}
\vspace{-.5cm}
\caption{LTS abstracting composition building procedure}
\label{lst:build-abstraction}
\vspace{-.5cm}
\end{figure}

Once we have built the \k{abstracting path graph}, obtaining a ranking for the actions enabled from the state $\<s_A,\ldots,s_N\>$ is straightforward. The states reached by the $\mk{goals}$ transitions are set with an estimated distance to a goal of $0$, while the states in the $\mk{errors}$ set are set with an $\infty$ estimated distance. Then, we propagate these estimates from the states to its parents following the $\mk{edges}$ backwards. For each step we increase the estimated distance between parent and child according to the information stored in the $\mk{generations}$ map. For a state with multiple children we keep as an estimate the minimum of its children's,
\insight{since we are interested in reaching the goal as fast as possible.}
When this back-propagation ends, all the enabled actions have an estimated distance to the goal, thus it is just a matter of sorting them with respect to these values. However, the same action could have multiple estimated values since we allow not only intra-LTS transitions but also inter-LTS transitions; in these cases we keep the best (minimum) estimate.

Summarizing, the procedure computes\insight{ -- through the construction of the graph --}
the length of the \k{abstracting paths} starting at a given initial state and reaching a co-safety goal. The algorithm concentrates on the best estimates, which are used to guide the on-the-fly exploration.
Thus, it effectively computes an admissible heuristic, since it can never overestimate the distance to the goal (Lemma~\ref{lem:admissibility}).

%%%%%%%%%%%%%%%%%%%%%%%%%%%%%%%%%%%%%%%%%%%%%%%%%%%%%%%%%%%%%%%%%%%%%%%%%%%%%%%
%%  Evaluation
%%%%%%%%%%%%%%%%%%%%%%%%%%%%%%%%%%%%%%%%%%%%%%%%%%%%%%%%%%%%%%%%%%%%%%%%%%%%%%%

\section{Evaluation} \label{sec:evaluation}

In this section we report on an evaluation of DCS. We compare DCS with three other approaches running on an Intel i7-3770 with 8GB of RAM:

\begin{enumerate}
\item Monolithic explicit state representation, previously implemented in MTSA\footnote{Available at https://bitbucket.org/dciolek/mtsa} (similar to CTCT \cite{Wonham:1999:NCDES}).
\item Monolithic symbolic state representation using BDD, implemented in MBP \cite{Gromyko:2006:TCS}.
\item Compositional explicit state representation implemented in Supremica (SUP) \cite{Mohajerani:2011:CSDES}.
\end{enumerate}

The aim of the evaluation is to assess the gains in scalability with the use of the informed search procedure. For this reason, we want case studies with the capability to scale up to higher complexities. Fortunately, both MBP and Supremica come bundled with the specification for one of the most traditional examples in controller synthesis: Line Transfer (LT); which was first introduced by Wonham \cite{Wonham:1999:NCDES}. The fact that the models were written by the tools authors' is important because it reduces the impact of a threat to validity that would be an ill-designed model. However, in order to be able to scale the problem up we had to extend the models, yet we maintain the same structure.

Given the complexity of: writing identical specifications in the different formal languages, finding examples that can be scaled up and presenting the results; we opt for concentrating in varying three independent parameters for the LT. This gives us a big enough number of environments with different topologies to evaluate the techniques. Still, this represents another threat to validity, since ideally the techniques should be compared with a more diverse set of inputs.

The LT consists of series of machines $M_1, M_2, \ldots, M_n$ connected by buffers $B_1, B_2, \ldots, B_n$ and ending in a special machine called Test Unit ($TU$). A machine $M_i$ takes work pieces from the buffer $B_{i-1}$ (with the exception of machine $M_1$ that takes the work pieces from the outside). After an undetermined amount of time, the working machine $M_i$ outputs a processed work piece through buffer $B_i$. Finally, when a work piece reaches the $TU$ it can be accepted and taken out of the system or it can be rejected and placed back in buffer $B_1$ for reprocessing. The only controllable actions in this case study are the taking of work pieces. An error ensues if a machine tries to take a work piece from an empty buffer or if it tries to place a processed work piece in a full buffer. One of the goals for the controller is to \k{avoid} the actions that lead to errors, the other goal is to \k{reach} a state where a processed work piece can be accepted or rejected. \insight{We do not require the controller to achieve accepted pieces as acceptance and rejection are not decided by the controller.}

The case study can be scaled in three directions:

\begin{enumerate}
\item $Machines$ (M): number of interconnected machines.
\item $Workload$ (W): maximum number of work pieces a machine can process simultaneously.
\item $Capacity$ (C): capacity of the buffers.
\end{enumerate}

\begin{example}
In Fig.~\ref{lst:stl-fsp} we present the TL model as accepted by MTSA, given in the Finite State Processes modeling language (FSP). Technically, FSP is a process calculus, in the spirit of CSP, designed to be easily machine and human readable. FSP includes standard constructs such as action prefix (\v{->}), external choice (\v{|}), alphabet extension (\v{+}) and parallel composition (\v{||}).

We model the \v{Machines} and \v{TU} separately; and afterwards the \v{Buffer}. The \v{Machine} starts idle and can \v{get} as many work pieces as its workload allows it. Uncontrollably the processing ends and the element is \v{put} in the next buffer. The \v{TU} gets an element from the last buffer and can either \v{accept} it or \v{reject} it making it \v{return} to the first buffer. Note that on attempting an invalid operation the \v{Buffer} goes into an \v{ERROR} state (i.e. a deadlock). The \v{||Plant} process represents the parallel composition between all the processes.

\begin{figure}
\vspace{0.2cm}
\begin{framed}
\begin{lstlisting}
Machine(Id=0) = Working[0],
  Working[w:0..W] =
    (when (w < W) get[Id]   -> Working[w+1] |
     when (w > 0) put[Id+1] -> Working[w-1] ).

TU = Idle,
  Idle    = (get[M] -> Testing ),
  Testing = (ret[1] -> reject -> Idle |
             accept -> Idle)
            +{ret[0..M]}.

Buffer(Id=0) = At[0],
  At[c:0..C] = (
    when (c > 0) get[Id] -> At[c-1] |
    when (c = 0) get[Id] -> ERROR   |
    when (c < C) put[Id] -> At[c+1] |
    when (c = C) put[Id] -> ERROR   |
    when (c < C) ret[Id] -> At[c+1] |
    when (c = C) ret[Id] -> ERROR   ).

||Plant = (forall [m:0..M-1] (
    Machine(m) || Buffer(m+1)) || TU).
\end{lstlisting}
\end{framed}
\vspace{-.5cm}
\caption{FSP model of the Transfer Line case study}
\label{lst:stl-fsp}
\vspace{-.25cm}
\end{figure}
\end{example}

In Table~\ref{tbl:small-scale} we consider some ``small scale'' cases, in which we consider all the combinations of the parameters for the following values: M in $[4,5,6]$, W in $[1,2,3]$, C in $[1,2,3]$. We report the time in seconds required by each tool to synthesize a controller with a Time Out (TO) of 30 minutes.

As it can be seen in the table, the number of states in the environment grows very quickly and it soon takes the tools to their limits. Nonetheless, DCS is able to solve all the problems under the second mark.
\insight{
MTSA soon runs Out of Memory (OM) while, despite also working with a full representation of the environment, MBP solves many problems including large instances. Whereas Supremica outperforms the other methods in the simpler cases but falls short on the more complex ones.}

\begin{table}[t]
\centering
\vspace{0.2cm}
\begin{scriptsize}
\begin{tabular}{|c|c|c|c||C{25pt}|C{25pt}|C{25pt}|C{25pt}|}
\hline 	M	&W	&C	&States			&MTSA 	&DCS     	&MBP 	&SUP     	\\
\hline
\hline	4	&1	&1	&\rb{1.5e$^4$}	&1.5 	&0.01    	&0.07	&\bf 0.01	\\
\hline	4	&1	&2	&\rb{3.5e$^4$}	&5.5 	&0.02    	&0.21 	&\bf 0.01	\\
\hline	4	&1	&3	&\rb{7.2e$^4$}	&29.8	&0.03    	&0.41 	&\bf 0.03	\\
\hline	4	&2	&1	&\rb{7.7e$^4$}	&9.1 	&0.01    	&0.11 	&\bf 0.01	\\
\hline	4	&2	&2	&\rb{1.8e$^5$}	&74.5	&\bf 0.01  	&0.86 	&0.04    	\\
\hline	4	&2	&3	&\rb{3.6e$^5$}	&OM		&\bf 0.01  	&5.95	&0.18    	\\
\hline	4	&3	&1	&\rb{2.3e$^5$}	&OM		&\bf 0.01  	&0.08	&0.02    	\\
\hline	4	&3	&2	&\rb{5.6e$^5$}	&OM		&\bf 0.01  	&0.87 	&0.03    	\\
\hline	4	&3	&3	&\rb{2.2e$^6$}	&OM		&\bf 0.02  	&11.58	&0.45    	\\
\hline	5	&1	&1	&\rb{1.2e$^5$}	&33.3	&0.01    	&0.14 	&\bf 0.01	\\
\hline	5	&1	&2	&\rb{3.5e$^5$}	&OM		&\bf 0.02  	&1.43	&0.09    	\\
\hline	5	&1	&3	&\rb{8.6e$^5$}	&OM		&\bf 0.05  	&3.53	&0.72    	\\
\hline	5	&2	&1	&\rb{8.8e$^5$}	&OM		&\bf 0.01  	&0.25 	&0.09    	\\
\hline	5	&2	&2	&\rb{2.6e$^6$}	&OM		&\bf 0.02  	&12.27	&0.62    	\\
\hline	5	&2	&3	&\rb{6.7e$^6$}	&OM		&\bf 0.03  	&118 	&62      	\\
\hline	5	&3	&1	&\rb{3.7e$^6$}	&OM		&\bf 0.01  	&0.26 	&0.22    	\\
\hline	5	&3	&2	&\rb{1.1e$^7$}	&OM		&\bf 0.02  	&13.28	&0.43    	\\
\hline	5	&3	&3	&\rb{2.8e$^7$}	&OM		&\bf 0.06  	&263.72	&163     	\\
\hline	6	&1	&1	&\rb{9.7e$^5$}	&OM		&0.22    	&0.47 	&\bf 0.09	\\
\hline	6	&1	&2	&\rb{3.5e$^6$}	&OM		&\bf 0.18  	&14.66	&0.77    	\\
\hline	6	&1	&3	&\rb{1.0e$^7$}	&OM		&\bf 0.21  	&37.91	&TO  		\\
\hline	6	&2	&1	&\rb{1.1e$^7$}	&OM		&\bf 0.02  	&1.75	&0.13    	\\
\hline	6	&2	&2	&\rb{4.0e$^7$}	&OM		&\bf 0.07 	&200 	&79      	\\
\hline	6	&2	&3	&\rb{1.2e$^8$}	&OM		&\bf 0.25  	&1397 	&TO   		\\
\hline	6	&3	&1	&\rb{5.9e$^7$}	&OM		&\bf 0.03  	&1.23	&0.17    	\\
\hline	6	&3	&2	&\rb{2.2e$^8$}	&OM		&\bf 0.04 	&228 	&177     	\\
\hline	6	&3	&3	&\rb{6.7e$^8$}	&OM		&\bf 0.22  	&TO		&TO   		\\
\hline
\end{tabular}
\end{scriptsize}
\caption{Small scale Line Transfer results}
\label{tbl:small-scale}
\vspace{-.5cm}
\end{table}

\begin{example}
Analyzing the controller generated by DCS for the TL case study, we see that the heuristic guides the search for the controller disregarding unnecessary actions such as simultaneously processing multiple work pieces. The controller found by DCS is roughly a sequence of states such that each state is connected with the following with interleaving $get$ and $put$ actions until an $accept$ or $reject$ becomes available. In Fig.\ref{fig:TLM2W1B1} we show the controller generated by DCS for $2$ machines. Despite the fact that the technique pursues a reachability objective, it accidentally finds controllers with cyclic behavior, this is caused by the reopening of already explored states. However, this is not to be expected in general.

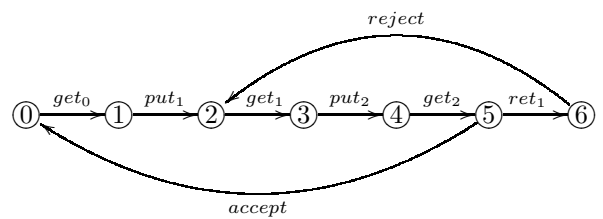
\begin{figure}[ht]
\centering
$
\xymatrix@+0pc{
\node{0} \ar^{get_0}[r] & \node{1} \ar^{put_1}[r] & \node{2} \ar^<<<<<<{get_1}[r] & \node{3} \ar^{put_2}[r] & \node{4} \ar^{get_2}[r] & \node{5} \ar^{accept}@/^2.5pc/[0,-5] \ar^<<<<{ret_1}[r] & \node{6} \ar_{reject}@/_2.5pc/[0,-4] \\
}$
\caption{Controller generated by DCS for a TL with $2$ machines} \label{fig:TLM2W1B1}
\end{figure}
\end{example}

%An advantage of the \k{LTS abstraction} is that it provides much more information than that used for the evaluation of a single state. This allows us to introduce a simple optimization that avoids unnecessary computation reusing a previously computed abstraction until a precise estimate cannot be obtained without rebuilding the abstraction. We have found that for this case the optimization significantly reduces the computational cost of the search procedure.

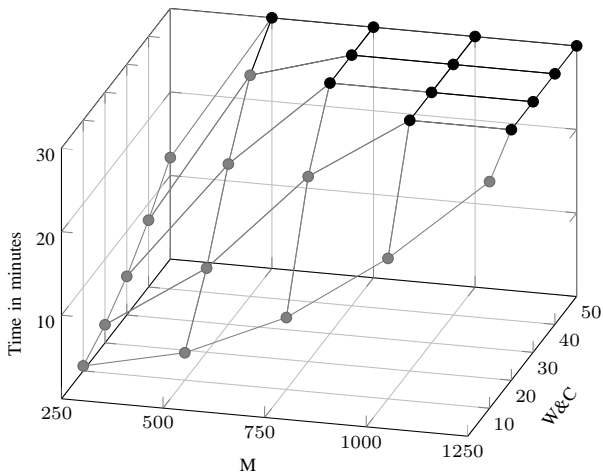
\begin{figure}
\pgfplotsset{
colormap={blackwhite}{[1pt]
  % white: from 0000 to 1700
  rgb(0000pt)=(0.5,0.5,0.5);
  rgb(1700pt)=(0.5,0.5,0.5);
  % black: from 1700 to 1800
  rgb(1700pt)=(0.0,0.0,0.0);
  rgb(1800pt)=(0.0,0.0,0.0);
},
}
\vspace{0.2cm}
\begin{scriptsize}
\begin{tikzpicture}
\pgfkeys{%
    /pgf/number format/set thousands separator = {}}
\begin{axis}[enlargelimits=false,ymin=0,zmin=0,grid=major,view={15}{30},
             xtick={250,500,750,1000,1250},ytick={10,20,30,40,50},ztick={600,1200,1800},zticklabels={10,20,30},
             xlabel={M},ylabel style={sloped},ylabel={W\&C},zlabel={Time in minutes}]

\addplot3[mesh,scatter]
coordinates{ 
(250,10,35)  (500,10,197)  (750,10,517)  (1000,10,1009) (1250,10,1628)

(250,20,130) (500,20,604)  (750,20,1329) (1000,20,1800) (1250,20,1800)

(250,30,277) (500,30,1151) (750,30,1800) (1000,30,1800) (1250,30,1800)

(250,40,480) (500,40,1588) (750,40,1800) (1000,40,1800) (1250,40,1800)

(250,50,729) (500,50,1800) (750,50,1800) (1000,50,1800) (1250,50,1800)
};

\end{axis}

%\begin{axis}

%\addplot3
%coordinates {
%  (250,10,35)  (500,10,197)  (750,10,517)  (1000,10,1009) (1250,10,1728)
%  (250,20,130) (500,20,604)  (750,20,1429) (1000,20,1800) (1250,20,1800)
%  (250,30,277) (500,30,1151) (750,30,1800) (1000,30,1800) (1250,30,1800)
%  (250,40,480) (500,40,1688) (750,40,1800) (1000,40,1800) (1250,40,1800)
%  (250,50,729) (500,50,1800) (750,50,1800) (1000,50,1800) (1250,50,1800)
%};
%coordinates {
%  (250,10,35)  (250,20,130)  (250,30,277)  (250,40,480)  (250,50,729)
%  (500,10,197) (500,20,604)  (500,30,1151) (500,40,1688) (500,50,1800)
%  (750,10,517) (750,20,1429) (750,30,1800) (750,40,1800) (750,50,1800)
%};
%\addplot3 
%coordinates { (250,10,35)  (500,10,197)  (750,10,517)  (1000,10,1009) (1250,10,1728) };
%\addplot3 
%coordinates { (250,20,130) (500,20,604)  (750,20,1429) (1000,20,1800) (1250,20,1800) };
%\addplot3 
%coordinates { (250,30,277) (500,30,1151) (750,30,1800) (1000,30,1800) (1250,30,1800) };
%\addplot3 
%coordinates { (250,40,480) (500,40,1688) (750,40,1800) (1000,40,1800) (1250,40,1800) };
%\addplot3 
%coordinates { (250,50,729) (500,50,1800) (750,50,1800) (1000,50,1800) (1250,50,1800) };

%\addplot3
%coordinates { (500,10,197) (500,20,604)  (500,30,1151) (500,40,1688) (500,50,1800) };
%\addplot3
%coordinates { (750,10,517) (750,20,1429) (750,30,1800) (750,40,1800) (750,50,1800) };

%\end{axis}

\end{tikzpicture}
\end{scriptsize}
\vspace{-.25cm}
\caption{Big scale LT results (TOs are depicted with black dots)}
\label{fig:big-scale}
\vspace{-.5cm}
\end{figure}

In a second experiment we assess the boundaries of the technique using ``big scale'' cases shown in Fig.~\ref{fig:big-scale}. We remove the other techniques from the evaluation since they cannot tackle these problems. For ease of presentation we report on a subset of the combinations of the parameters where W and C take the same values.
\insight{Note that W and C are directly related with the number of states in the Machine and Buffer LTSs respectively, while M relates to the number of components. Thus, the evaluation still allows to see how the technique performs as the complexity and the number of components increase independently.}

\insight{Summarizing, we have evaluated DCS against other approaches to controller synthesis. DCS shows promise given that it is able to cope with increasing complexities beyond the capabilities of similar tools. The largest case tackled by DCS considers an environment with $8.9$e$^{2734}\!$ states (solved in $1728s$) generating a controller with $2503$ states. }

%%%%%%%%%%%%%%%%%%%%%%%%%%%%%%%%%%%%%%%%%%%%%%%%%%%%%%%%%%%%%%%%%%%%%%%%%%%%%%%
%%  Conclusions
%%%%%%%%%%%%%%%%%%%%%%%%%%%%%%%%%%%%%%%%%%%%%%%%%%%%%%%%%%%%%%%%%%%%%%%%%%%%%%%

\section{Conclusions and future work} \label{sec:conclusions}

In this paper we present DCS, a method that looks for a controller exploring the state space on-the-fly guided by a domain-independent heuristic. The method performs better than other approaches both in time and memory for the selected case study, despite its worst case complexity. The approach forgoes maximality since reachability goals allow for a directed search in a reduced portion of the state space.

The risk of DCS is that it could fail to properly guide the exploration, triggering an unnecessary large exploration of the state space. Thus, despite our efforts to keep the approach as general as possible, further experiments are required in order to properly asses the generality of the technique. For this reason we plan to work on automatic translations to simplify the comparison with other tools.

These preliminary results show promise and hence we intend to extend the technique. Our primary concern is to deal with general liveness goals, since the domain of application of controller synthesis usually have these kind of requirements.

%since controller synthesis problems usually have these kind of requirements.
%since the domain of application of controller synthesis usually have these kind of requirements.

We highlight that a compact representation of the problem is essential for extracting useful guidance for the informed search procedure. Thus, when dealing with DES, elements such as components and synchronization provide vital information to take into account. Furthermore, even when the \k{LTS abstraction} achieves remarkable results, there are other heuristics that could be used (instead or in combination) that could perform better. The field of directed controller synthesis is almost uncharted, and we believe it can lead to a leap in the applicability of controller synthesis techniques.

%%%%%%%%%%%%%%%%%%%%%%%%%%%%%%%%%%%%%%%%%%%%%%%%%%%%%%%%%%%%%%%%%%%%%%%%%%%%%%%
%%  Bibliography
%%%%%%%%%%%%%%%%%%%%%%%%%%%%%%%%%%%%%%%%%%%%%%%%%%%%%%%%%%%%%%%%%%%%%%%%%%%%%%%
\bibliographystyle{ieeetr}
\bibliography{bibliography}

\end{document}